\documentclass[a4paper,11pt]{article}
\pdfoutput=1
\usepackage{jcappub}

\usepackage[utf8]{inputenc}

\usepackage{amsmath, amssymb}
\usepackage{graphicx, xcolor}
\usepackage{epsfig}
\usepackage{latexsym}
\usepackage{xspace}
\usepackage{enumerate}
\usepackage{color}
\usepackage{mathtools}

\newcommand{\rR}{\rho_R}
\newcommand{\rbh}{\rho_\text{BH}}
\newcommand{\gs}{g_\star}
\newcommand{\gss}{g_{\star s}}
\newcommand{\Teq}{T_\text{eq}}
\newcommand{\Min}{M_\text{in}}
\newcommand{\Tin}{T_\text{in}}
\newcommand{\tin}{t_\text{in}}
\newcommand{\Tev}{T_\text{ev}}
\newcommand{\tev}{t_\text{ev}}
\newcommand{\Tc}{T_\text{c}}
\newcommand{\Mbh}{M_\text{BH}}
\newcommand{\Tbh}{T_\text{BH}}
\newcommand{\Tosc}{T_\text{osc}}
\newcommand{\Tqcd}{T_\text{QCD}}

\title{Axion Dark Matter\\in the Time of Primordial Black Holes}
\author[a]{Nicolás Bernal,}
\author[b,c]{Fazlollah Hajkarim}
\author[d]{and Yong Xu}
\affiliation[a]{Centro de Investigaciones, Universidad Antonio Nariño\\Carrera 3 este \# 47A-15, Bogotá, Colombia}
\affiliation[b]{Dipartimento di Fisica e Astronomia, Universit\`a degli Studi di Padova \\Via Marzolo 8, 35131 Padova, Italy  }
\affiliation[c]{
Istituto Nazionale di Fisica Nucleare (INFN), Sezione di Padova\\Via Marzolo 8, 35131 Padova, Italy}
\affiliation[d]{Bethe Center for Theoretical Physics and Physikalisches Institut, Universit\"at Bonn\\Nussallee 12, D-53115 Bonn, Germany}

\emailAdd{nicolas.bernal@uan.edu.co}
\emailAdd{hajkarim@pd.infn.it}
\emailAdd{yongxu@th.physik.uni-bonn.de}

\abstract{We investigate the production of QCD axion dark matter in a nonstandard cosmological era triggered by primordial black holes (PBHs) that fully evaporate before the onset of BBN. Even if PBHs cannot emit the whole axion cold dark matter abundance through Hawking radiation, they can have a strong impact on the dark matter produced via the misalignment mechanism. First, the oscillation temperature of axions reduces if there is a PBH dominated era, and second, PBH evaporation injects entropy to the standard model, diluting the axion relic abundance originally produced.
The axion window is therefore enlarged, reaching masses as light as $\sim 10^{-8}$~eV and decay constants as large as $f_{a}\sim 10^{14}$~GeV without fine tuning the misalignment angle.
Such small masses are in the reach of future detectors as ABRACADABRA, KLASH, and ADMX, if the axion couples to photons.
Additionally, the axions radiated by PBHs contribute to $\Delta N_\text{eff}$ within the projected reach of the future CMB Stage 4 experiment.}
\begin{document}
\begin{flushright}
  PI/UAN-2021-694FT
\end{flushright}

\maketitle

%%%%%%%%%%%%%%%%%%%%%%%%%%%%%%%%%%%%%%%%%%%%%%%%%%%%%%%%%%
\section{Introduction}
%%%%%%%%%%%%%%%%%%%%%%%%%%%%%%%%%%%%%%%%%%%%%%%%%%%%%%%%%%
QCD axion is a  pseudo-Nambu-Goldstone boson arising from the spontaneous breaking of a global Peccei-Quinn symmetry~\cite{Peccei:1977hh}.
At high temperature axion is massless, however, a mass and potential can be generated due to the non-perturbative QCD effect during  quark condensation~\cite{Wilczek:1977pj, Weinberg:1977ma}. The axion oscillates around the minimum of its potential, which dynamically generates a CP conserved phase and predicts a vanishing neutron electric dipole moment, naturally solving the Strong CP problem.  Another interesting outcome from this (misalignment) process is that axion can account for the whole cold dark matter (DM) in the universe~\cite{Preskill:1982cy, Abbott:1982af, Dine:1982ah}. The ``two birds with one stone" feature of axion has been attracting many investigations, for reviews see, e.g., Refs.~\cite{Marsh:2015xka, DiLuzio:2020wdo, Sikivie:2020zpn}.

In standard cosmology, the predicted axion relic density via the misalignment mechanism depends on the axion mass and the initial misalignment angle (expected to be $\mathcal{O}(1)$). This leads to a narrow range of axion mass $m_{a} \simeq 10^{-6}$~eV (or correspondingly $f_a \simeq 10^{12}$~GeV) to fit the correct DM relic density.%
\footnote{Another possible source for axion production is the decay of topological defects. It gives rise to a different window for the required axion decay constant to obtain the correct relic density~\cite{Sikivie:1982qv, DiLuzio:2020wdo}.}
However, scenarios yielding nonstandard cosmologies widen the axion window~\cite{Steinhardt:1983ia, Lazarides:1990xp, Kawasaki:1995vt, Giudice:2000ex, Grin:2007yg, Visinelli:2009kt, Nelson:2018via, Visinelli:2018wza, Ramberg:2019dgi, Blinov:2019jqc, Allahverdi:2020bys, Carenza:2021ebx, Heurtier:2021rko, Venegas:2021wwm, Arias:2021rer}. 
For example, lighter axions with masses $m_a \gtrsim \mathcal{O}(10^{-8})$~eV can appear in scenarios featuring an early matter dominated phase, whereas heavier axions $m_a \lesssim \mathcal{O}(10^{-2})$~eV can be produced if the universe had a kination epoch~\cite{Arias:2021rer}.

Alternatively, a nonstandard epoch could have been triggered by primordial black holes (PBHs), which might have been copiously  formed  due to inhomogeneties  of density fluctuations in the early universe~\cite{Zeldovich:1967lct, Carr:1974nx, Dolgov:1992pu},
see e.g. Refs.~\cite{Sasaki:2018dmp, Carr:2020xqk} for reviews. The  existence of PBHs is nowadays well  supported  in light of the recent observations, for more detailed discussions, see e.g., Refs.~\cite{Bird:2016dcv, Sasaki:2016jop, Clesse:2016vqa, Clesse:2017bsw, Gow:2019pok}. PBHs after formation behave as matter, which could have constituted a large fraction of the energy budget of the universe.

PBHs, and particularly PBH dominated eras, can have a number of phenomenological consequences in the early universe. For instance, they can trigger baryogenesis~\cite{Barrow:1990he,Hamada:2016jnq,Hooper:2020otu, Datta:2020bht}, or radiate DM or dark radiation via their Hawking evaporation~\cite{Hooper:2019gtx, Masina:2020xhk, Gondolo:2020uqv, Baldes:2020nuv, Bernal:2020kse, Bernal:2020ili, Bernal:2020bjf, Cheek:2021odj, Cheek:2021cfe, JyotiDas:2021shi, Arbey:2021ysg, Masina:2021zpu}, or even  source DM, baryogenesis and  density perturbations simultaneously~\cite{Fujita:2014hha}.  Moreover, they may  also play an important role in  neutrino physics, like sourcing massive neutrinos~\cite{Lunardini:2019zob, Perez-Gonzalez:2020vnz} or leaving a signal on the neutrino floor~\cite{Calabrese:2021zfq}. Finally, axion-like particles generated from PBHs may also lead to some observable signals on the cosmic $X$-ray background~\cite{Schiavone:2021imu}.

Different from the nonstandard scenarios considered in Refs.~\cite{Steinhardt:1983ia, Lazarides:1990xp, Kawasaki:1995vt, Giudice:2000ex, Grin:2007yg, Visinelli:2009kt, Nelson:2018via, Visinelli:2018wza, Ramberg:2019dgi, Blinov:2019jqc, Allahverdi:2020bys, Carenza:2021ebx, Arias:2021rer} and along  the recent line of study  with an early PBHs phase  mentioned above,  we investigate the phenomenological consequences of the  axion as DM with a nonstandard  epoch  triggered PBHs.  
We should note that there is a difference between a period of matter domination by a heavy field field and by PBHs. The latter one behaves like a decaying field with a time-dependent decay rate. This makes the first and second order gravitational wave signatures (i.e., the slope of the spectrum) of matter dominated era via a heavy field and PBHs potentially distinguishable~\cite{Nakayama:2008wy, Bernal:2019lpc, Ramberg:2019dgi, Papanikolaou:2020qtd, Domenech:2020ssp}.
We focus on light PBHs, with masses smaller than $\sim 2\times 10^8$~g, which fully evaporate before the onset of Big Bang nucleosynthesis (BBN). PBHs modify the standard axion production in different ways: $i)$ axions are produced inevitably from Hawking radiation by PBH evaporation, $ii)$ the oscillation temperature of axions reduces if there is a PBH dominated era, and $iii)$ PBH evaporation injects entropy to the SM, diluting the axion relic abundance originally produced by the misalignment mechanism. 

We find that the overall effect is that the parameter space with ultralight axions is opened up due  to  the entropy injection from PBH  evaporation. In particular, the lower limit for axion DM mass  can reach to $m_a \sim \mathcal{O}(10^{-8})$~eV (or correspondingly $f_a \sim 10^{14}$~GeV) even without fine tuning the  misalignment angle. Interestingly, such a small mass of this order is within the reach of future detectors such as ABRACADABRA~\cite{Kahn:2016aff, Ouellet:2018beu}, KLASH~\cite{Alesini:2017ifp, Alesini:2019nzq}, and the next generation of ADMX~\cite{ADMX:2009iij, ADMX:2019uok}, if the axion couples to photons.
Additionally, axions radiated by PBHs contribute to $\Delta N_\text{eff} \simeq 0.04$, within the projected reach of the future CMB Stage 4 experiment, and could relax the tension between late and early-time Hubble determinations.

The reminder of this paper is as follows.  We first revisit the axion DM in standard cosmology in Sec.~\ref{standardcos}. Then in Sec.~\ref{pbh} we set up the formalism for PBH evaporation and  analytically compute the entropy injection factor. In Sec.~\ref{axion_pbh_eva} we focus on the direct axion production channel from PBH evaporation and its contribution to the dark radiation. In Sec.~\ref{axion_PBH} we investigate effect of PBHs domination on axion DM abundance generated via misalignment mechanism. Finally, we summarize our results in Sec.~\ref{conclusion}. 

%%%%%%%%%%%%%%%%%%%%%%%%%%%%%%%%%%%%%%%%%%%%%%%%%%%%%%%%%%
\section{Axion DM in Standard Cosmology} \label{standardcos}
%%%%%%%%%%%%%%%%%%%%%%%%%%%%%%%%%%%%%%%%%%%%%%%%%%%%%%%%%%
Axion mass $m_a$ at zero temperature is given by~\cite{DiLuzio:2020wdo}
\begin{equation}
    m_a \simeq 5.7 \times 10^{-6} \left(\frac{10^{12}~\text{GeV}}{f_a}\right) \text{eV}\,,
\end{equation}
where $f_a$ denotes the decay constant. And the temperature-dependent axion mass $\tilde m_a$ is shown to be~\cite{Borsanyi:2016ksw}
\begin{equation}
    \tilde m_a(T) \simeq m_a \times
    \begin{dcases}
	    (\Tqcd/T)^4 & \text{for } T \geq \Tqcd\,,\\
	    1 & \text{for } T \leq \Tqcd\,,
    \end{dcases}
\end{equation}
with $\Tqcd \simeq 150$~MeV.

Axion begins to oscillate at the temperature $T = \Tosc$ defined by $3\,H(\Tosc) \equiv \tilde m_a(\Tosc)$, where $H(T) = \sqrt{\rR(T)/(3\,M_P^2)}$  denoting the Hubble expansion rate and 
\begin{equation}
    \rR(T) = \frac{\pi^2}{30}\, \gs(T)\, T^4
\end{equation}
is the SM radiation energy density and $\gs(T)$ corresponds to the number of relativistic degrees of freedom contributing to $\rR$.
Considering the conservation of the axion number density and assuming conservation of SM entropy, the energy density for non-relativistic axions $\rho_a$ at present is given by
\begin{equation} \label{rho0}
    \rho_a(T_0) = \rho_a(\Tosc) \frac{m_a}{\tilde m_a(\Tosc)} \frac{s(T_0)}{s(\Tosc)}\,,
\end{equation}
with $T_0$ the temperature today.  The SM entropy density is defined as 
\begin{equation}
    s(T) = \frac{2\pi^2}{45}\, \gss(T)\, T^3\,,
\end{equation}
where $\gss(T)$ denotes the corresponding  number of relativistic degrees \cite{Drees:2015exa}. 
Within the WKB approximation $\rho_a(\Tosc) \simeq \frac12 \tilde m_a^2(\Tosc)\, f_a^2\, \theta_i^2$, where $\theta_i$ is the initial misalignment angle~\cite{Hertzberg:2008wr, DiLuzio:2020wdo}.

\begin{figure}
    \def\sepf{0.57}
	\centering
    \includegraphics[scale=\sepf]{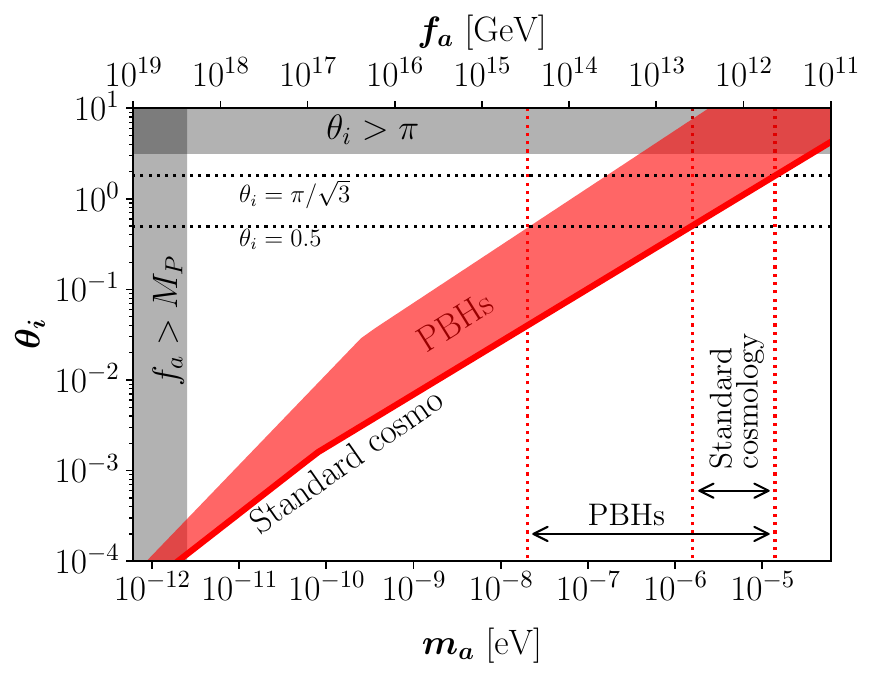}
    \caption{Misalignment angle required in order to reproduce the whole observed DM abundance for the standard cosmological case (solid red line) and the PBH domination (red band).
    In the gray band $f_a > M_P$ or $\theta_i > \pi$.
    }
	\label{fig:ma_theta}
\end{figure}

Using Eq.~\eqref{rho0}, the axion abundance is shown to be
\begin{equation}
    \begin{aligned}
        \Omega_ah^2 & \equiv \frac{\rho_a(T_0)}{\rho_c/h^2} \simeq 0.12 \left(\frac{\theta_i}{10^{-3}}\right)^2\\
        & \times
        \begin{dcases}
            \left(\frac{m_a}{m_a^\text{QCD}}\right)^{-\frac32} & \text{for } m_a \leq m_a^\text{QCD},\\
            \left(\frac{m_a}{m_a^\text{QCD}}\right)^{-\frac76} & \text{for } m_a \geq m_a^\text{QCD},
        \end{dcases}
    \end{aligned}
\end{equation}
with $m_a^\text{QCD} \equiv m_a(\Tosc=\Tqcd) \simeq 4.8 \times 10^{-11}$~eV, and where $\rho_c/h^2 \simeq 1.1 \times 10^{-5}$~GeV/cm$^3$ is the critical energy density and $s(T_0) \simeq 2.69 \times 10^3$~cm$^{-3}$ is the entropy density at present~\cite{Planck:2018vyg}.
The misalignment angle required to match the whole observed DM relic abundance (i.e., $\Omega_ah^2 \simeq 0.12$~\cite{Planck:2018vyg}) is shown with a thick red line in Fig.~\ref{fig:ma_theta}.
If $0.5 < \theta_i < \pi/\sqrt{3}$, $1.6 \times 10^{-6}~$eV $\lesssim m_a \lesssim 1.4 \times 10^{-5}$~eV, then it  corresponds to the usual QCD axion window in the standard cosmological scenario.
%%%%%%%%%%%%%%%%%%%%%%%%%%%%%%%%%%%%%%%%%%%%%%%%%%%%%%
%%%%%%%%%%%%%%%%%%%%%%%%%%%%%%%%%%%%%%%%%%%%%%%%%%%%%%

%%%%%%%%%%%%%%%%%%%%%%%%%%%%%%%%%%%%%%%%%%%%%%%%%%%%%%%%%%
%%%%%%%%%%%%%%%%%%%%%%%%%%%%%%%%%%%%%%%%%%%%%%%%%%%%%%%%%%
\section{Primordial Black Holes Evaporation and Entropy Injection} \label{pbh}
%%%%%%%%%%%%%%%%%%%%%%%%%%%%%%%%%%%%%%%%%%%%%%%%%%%%%%%%%%
PBHs could have been formed in a radiation dominated era, when the SM plasma has a temperature $T=\Tin$, with an initial mass $\Min$ similar to the enclosed mass in the particle horizon  given, and is given by~\cite{Carr:2009jm, Carr:2020gox}
\begin{equation}
    \Min \equiv \Mbh(\Tin) = \frac{4\pi}{3}\, \gamma\, \frac{\rR(\Tin)}{H^3(\Tin)}\,,
\end{equation}
with $\gamma \simeq 0.2$.
Extended PBH mass functions could arise naturally if the PBHs are generated from inflationary density fluctuations or cosmological phase transitions,  for a review, see e.g. Refs.~\cite{Sasaki:2018dmp, Carr:2020xqk}.
For the sake of simplicity  we have assumed that all PBHs have the same mass, (i.e., they were produced at the same temperature), which is a usual assumption in the literature.
Note also that PBHs can gain mass via mergers~\cite{Zagorac:2019ekv, Hooper:2019gtx, Hooper:2020evu} and accretion~\cite{Bondi:1952ni, Nayak:2009wk, Masina:2020xhk}.
However these processes are shown to be typically not very efficient, and could only induce mass gain of order $\mathcal{O}(1)$ \cite{Masina:2020xhk}, we will hereafter ignore them.

Particles lighter than BH horizon temperature~\cite{Hawking:1975vcx}
\begin{equation}
    \Tbh=\frac{M_P^2}{\Mbh}\simeq10^{13}\,{\rm GeV}\left(\frac{1\,{\rm g}}{\Mbh}\right)
\end{equation}
can be emitted via Hawking radiation, which can be described as a blackbody (up to greybody factors).  The energy spectrum of a species $j$ with $g_j$ internal degrees of freedom radiated by a nonrotating BH with zero charge can be described as~\cite{Page:1976df, Gondolo:2020uqv}
\begin{equation}
    \frac{d^2u_j(E,t)}{dt\,dE} = \frac{g_j}{8\pi^2} \frac{E^3}{e^{E/\Tbh}\pm 1}\,,
\end{equation}
where $+$ for fermions, $-$ for bosons, and  $u_j$ denotes the total radiated energy per unit area, $t$ the time, $E$ the energy of the emitted particle.

Evolution of the BH mass due to Hawking evaporation can be described as \cite{Gondolo:2020uqv}
\begin{equation} \label{eq:dMdt}
    \frac{d\Mbh}{dt} = -4\pi\,r_S^2\sum_j\int_0^\infty\frac{d^2u_j(E,t)}{dt\,dE}dE = -\frac{\pi\gs(\Tbh)}{480}\frac{M_P^4}{\Mbh^2}\,,
\end{equation}
where $r_S\equiv\frac{\Mbh}{4\pi\,M_P^2}$ is the Schwarzschild radius of the BH.
Neglecting the temperature dependence of $\gs$ during the whole lifetime of the BH, Eq.~\eqref{eq:dMdt} admits the analytical solution
\begin{equation} \label{massmin}
    \Mbh(t)=\Min\left(1-\frac{t-\tin}{\tau}\right)^{1/3},
\end{equation}
where $\tin$ corresponds to the time at formation, and
\begin{equation}
    \tau \equiv \frac{160}{\pi\,\gs(\Tin)}\frac{\Min^3}{M_P^4}
\end{equation}
is the PBH lifetime.

The initial PBH energy density is usually normalized to the SM energy density at formation via the parameter
\begin{equation}
    \beta \equiv \frac{\rbh(\Tin)}{\rR(\Tin) }\,.
\end{equation}
An early PBH-dominated era naturally happens eventually when $\rbh > \rR$, which corresponds to $\beta > \beta_c$, with
\begin{equation}
    \beta_c \equiv \frac{\Tev}{\Tin}\,,
\end{equation}
where $\Tev$ is the SM temperature at which PBHs completely evaporate.

It is worth noticing that the production of gravitational waves (GW) induced by small-scale density perturbations underlain by PBHs could lead to a backreaction problem~\cite{Papanikolaou:2020qtd}.
Additionally, stronger constraints on the amount of produced GWs comes from BBN~\cite{Domenech:2020ssp}.
However, this can be avoided if 
\begin{equation} \label{eq:GW}
    \beta \lesssim 3.3 \times 10^{-8} \left(\frac{\gamma}{0.2}\right)^{-\frac12}\left(\frac{\gs(T_{\text{BH}})}{108}\right)^{\frac{7}{16}}\left(\frac{\gs(T_{\text{ev}})}{106.75}\right)^{\frac{1}{16}} \left(\frac{\Min}{10^4~\text{g}}\right)^{-\frac78}.
\end{equation}
The evolution of PBH energy density and the SM entropy density can be tracked via the Boltzmann equations~\cite{Masina:2020xhk}:
\begin{align}
    \frac{d\rbh}{dt} + 3\, H\,\rbh =& +\frac{\rbh}{\Mbh}\, \frac{d\Mbh}{dt}\,, \label{boltz1}\\ 
    \frac{d\rR}{dt} + 4\, H\,\rR =& -\frac{\rbh}{\Mbh}\, \frac{d\Mbh}{dt}\,,  \label{boltz2}
\end{align}
where $H^2=(\rR+\rbh)/(3M_P^2)$.
We note that the SM entropy density is {\it not} conserved when PBHs evaporate.
As previously mentioned, for $\beta > \beta_c$ a matter domination induced by PBHs starts at $T = \Teq$, defined as $\rR(\Teq) \equiv \rbh(\Teq)$, with
\begin{equation}
    \Teq = \beta\, \Tin \left(\frac{\gss(\Tin)}{\gss(\Teq)}\right)^{1/3},
\end{equation}
and ends when PBHs fully evaporate, at $t = \tev = \tin + \tau \simeq \tau$, corresponding to a temperature 
\begin{equation} \label{Temtev}
    \Tev \simeq \left(\frac{\gs(\Tin)}{640}\right)^\frac14 \left(\frac{M_P^5}{\Min^3}\right)^\frac12\,.
\end{equation}
Requiring a successful BBN, one needs $\Tev \gtrsim 4$~MeV~\cite{Kawasaki:1999na, Kawasaki:2000en, deSalas:2015glj, Hasegawa:2019jsa}, which is equivalent to $\Min \lesssim 2 \times 10^8$~g.

Since PBHs radiate SM particles and therefore inject entropy to the SM bath, the SM radiation energy density would not scale as free radiation, but rather as $\rR(a) \propto a^{-3/2}$ if PBH dominates  the energy density and evolution of SM radiation, with $a$ being the scale factor. With this in mind, one can  characterize the expansion history into four distinct regimes~\cite{Arias:2019uol, Arias:2021rer}. The corresponding Hubble expansion rates are given by
\begin{equation} \label{eq:H}
    H(T) \simeq
    \begin{dcases}
        H_R(T) & \text{for } T \geq \Teq\,,\\
        H_R(\Teq) \left[\frac{\gss(T)}{\gss(\Teq)} \left(\frac{T}{\Teq}\right)^3\right]^{1/2} & \text{for } \Teq \geq T \geq \Tc\,,\\
        H_R(\Tev) \left[1 - \frac{720}{\pi\, \gs(\Tin)} \frac{\Min^3}{M_P^4} \frac{H_R^2(\Tev) - H_R^2(T)}{H_R(\Tev)}\right] & \text{for }  \Tc \geq T \geq \Tev\,,\\
        H_R(T) & \text{for } \Tev \geq T\,,
    \end{dcases}
\end{equation}
where
\begin{equation} \label{eq:Tc}
    \Tc \simeq \left[\frac{\gs(\Tin)\, \pi}{5760}\, \frac{M_P^{10}\, \Teq}{\Min^6}\right]^{1/5}
\end{equation}
corresponds to the temperature at which the evolution of the SM energy density starts to be dominated by the entropy injection of the PBHs.\footnote{Equation~\eqref{eq:Tc} can be obtained by matching the second and third lines of Eq.~\eqref{eq:H} at $\Tc$, and taking into account that $H_R(\Tc)$ dominates over $H_R(\Tev)$. In Appendix~\ref{appa} the detailed derivation for the third line of Eq.~\eqref{eq:H} is presented.}
By using Eq.\eqref{eq:H} one can show that the entropy injection is\footnote{In the Appendix the detailed calculation for the first line of Eq.~\eqref{eq:sratio} is shown.}
\begin{equation} \label{eq:sratio}
    \frac{S(T)}{S(\Tev)} \simeq
    \begin{dcases}
        \frac{\gss(\Teq)}{\gss(\Tev)}\, \frac{\gs(\Tev)}{\gs(\Teq)}\, \frac{\Tev}{\Teq} & \text{for } T \geq \Tc\,,\\
        \frac{\gss(T)}{\gss(\Tev)} \left(\frac{T}{\Tev}\right)^3 \left[1 - \frac{720}{\pi\, \gs(\Tin)} \frac{\Min^3}{M_P^4} \frac{H_R^2(\Tev) - H_R^2(T)}{H_R(\Tev)}\right]^{-2} & \text{for } \Tc \geq T \geq \Tev\,,\\
        1 & \text{for } \Tev \geq T\,.
    \end{dcases}
\end{equation}

A couple of comments are in order. In the regime $\Teq > T > \Tev$, PBH energy density dominates over radiation, giving rise to a nonstandard expansion rate of the universe, as shown in Eq.~\eqref{eq:H}. Additionally, within $\Tc > T > \Tev$, entropy injection from PBH evaporation dominates the evolution of the SM entropy density, and therefore SM radiation scales as $T(a) \propto a^{-3/8}$ instead of the usual $T(a) \propto a^{-1}$  characteristic of free radiation.%
\footnote{This comes from  the fact that during this period $\rR(a) \propto a^{-3/2}$ (cf. Eq.~\eqref{eq:sol_rhoR}), and that by definition $\rho_{\rm R}(T) \propto T^4$.}
Furthermore, when temperature $T > \Tev$ there is no SM entropy conservation, as seen in Eq.~\eqref{eq:sratio}. 
Finally, for $T < \Tev$, the standard cosmological scenario is recovered.

%%%%%%%%%%%%%%%%%%%%%%%%%%%%%%%%%%%%%%%%%%%%%%%%%%%%
\section{Axion Dark Radiation from PBH Evaporation} \label{axion_pbh_eva}
%%%%%%%%%%%%%%%%%%%%%%%%%%%%%%%%%%%%%%%%%%%%%%%%%%%%
PBHs inevitably radiate axions during their Hawking evaporation.
These axions are relativistic and therefore contribute to the universe's total energy density, potentially appearing as dark radiation.
Its contribution to the effective number of neutrinos $\Delta N_\text{eff} \simeq 0.04$~\cite{Hooper:2019gtx, Schiavone:2021imu} is below current constraints from measurements of the cosmic microwave background (CMB) and baryon acoustic oscillations~\cite{Planck:2018vyg}, but within the projected reach of the future CMB Stage~4 experiment~\cite{CMB-S4:2016ple, NASAPICO:2019thw}.
Additionally, it is interesting to note that this contribution is suited to relax the tension between the value of the Hubble constant as determined from local measurements~\cite{Riess:2016jrr, DEramo:2018vss, Riess:2018byc, Riess:2019cxk} and as inferred from the temperature anisotropies of the CMB~\cite{Planck:2018vyg}.

%%%%%%%%%%%%%%%%%%%%%%%%%%%%%%%%%%%%%%%%%%%%%%%%%%%%%%%
%%%%%%%%%%%%%%%%%%%%%%%%%%%%%%%%%%%%%%%%%%%%%%%%%%%%%%%
\section{Axion DM in the time of PBHs} \label{axion_PBH}
Even if PBHs are not able to radiate the whole axion DM abundance, they could have a strong impact on its genesis, due to a  non-standard cosmological era triggered if $\beta > \beta_c$.
Their effect is twofold: $i)$ an enhancement of the Hubble expansion rate due to the induction of an early matter dominated era, and therefore a reduction of the temperature at which axion starts to oscillate. And $ii)$ a dilution of the DM abundance by the entropy injection produced by the PBH evaporation.
In this case, taking into account the entropy injection, the axion energy density at present becomes
\begin{equation} \label{eq:rhoa}
    \rho_a(T_0) = \rho_a(\Tosc) \frac{m_a}{\tilde m_a(\Tosc)} \frac{s(T_0)}{s(\Tosc)} \times \frac{S(\Tosc)}{S(T_\text{ev})}\,.
\end{equation}
The modification of the Hubble expansion rate in Eq.~\eqref{eq:H} induces a decrease of $\Tosc$. 
For $\Tosc \leq \Tqcd$,
\begin{equation} \label{eq:Tosc1}
    \Tosc \simeq
    \begin{dcases}
        \left(\frac{1}{\pi} \sqrt{\frac{10}{\gs}}\, M_P\, m_a\right)^\frac12 & \text{for } T \geq \Teq\,,\\
        \left(\frac{1}{\pi} \sqrt{\frac{10}{\gs}}\, \frac{M_P\, m_a}{\sqrt{\Teq}}\right)^\frac23 & \text{for } \Teq \geq T \geq \Tc\,,\\
        \left(\frac{1}{\pi} \sqrt{\frac{10}{\gs}}\, M_P\, m_a\, \Tev^2\right)^\frac14 & \text{for }  \Tc \geq T \geq \Tev\,,\\
        \left(\frac{1}{\pi} \sqrt{\frac{10}{\gs}}\, M_P\, m_a\right)^\frac12 & \text{for } \Tev \geq T\,.
    \end{dcases}
\end{equation}
Alternatively, for $\Tosc \geq \Tqcd$,
\begin{equation} \label{eq:Tosc2}
    \Tosc \simeq
    \begin{dcases}
        \left(\frac{10}{\gs\, \pi^2}\, \Tqcd^8\, M_P^2\, m_a^2\right)^\frac{1}{12} & \text{for } T \geq \Teq\,,\\
        \left(\frac{10}{\gs\, \pi^2}\, \frac{\Tqcd^8\, M_P^2\, m_a^2}{\Teq}\right)^\frac{1}{11} & \text{for } \Teq \geq T \geq \Tc\,,\\
        \left(\frac{10}{\gs \pi^2} \Tqcd^8 M_P^2\, m_a^2\, \Tev^4\right)^\frac{1}{16} & \text{for }  \Tc \geq T \geq \Tev,\\
        \left(\frac{10}{\gs\, \pi^2}\, \Tqcd^8\, M_P^2\, m_a^2\right)^\frac{1}{12} & \text{for } \Tev \geq T\,.
    \end{dcases}
\end{equation}

Equations~\eqref{eq:rhoa}-\eqref{eq:Tosc2} together with Eq.~\eqref{eq:sratio} allow to estimate the axion DM abundance, which in the case $\Tosc \geq \Tqcd$ becomes
\begin{equation} \label{eq:DM}
    \frac{\Omega_a h^2}{0.12} \simeq
    \begin{dcases}
        \left(\frac{\theta_i}{1}\right)^2 \left(\frac{m_a}{10^{-7}~\text{eV}}\right)^{-\frac76} \left(\frac{\beta}{2 \times 10^{-12}}\right)^{-1} \left(\frac{\Min}{10^8~\text{g}}\right)^{-1} & \text{for } \Tosc \geq \Teq\,,\\
        \left(\frac{\theta_i}{1}\right)^2 \left(\frac{m_a}{10^{-7}~\text{eV}}\right)^{-\frac{14}{11}} \left(\frac{\beta}{10^{-13}}\right)^{-\frac{4}{11}} \left(\frac{\Min}{3 \times 10^8~\text{g}}\right)^{-\frac{29}{22}} & \text{for } \Teq \geq \Tosc \geq \Tc\,,\\
        \left(\frac{\theta_i}{1}\right)^2 \left(\frac{m_a}{10^{-8}~\text{eV}}\right)^{-\frac32} \left(\frac{\Min}{10^8~\text{g}}\right)^{-3} & \text{for } \Tc \geq \Tosc \geq \Tev\,,\\
        \left(\frac{\theta_i}{10^{-3}}\right)^2 \left(\frac{m_a}{m_a^\text{QCD}}\right)^{-\frac76} & \text{for } \Tev \geq \Tosc\,,
    \end{dcases}
\end{equation}
whereas in the case $\Tosc \leq \Tqcd$,
\begin{equation}
    \frac{\Omega_a h^2}{0.12} \simeq
    \begin{dcases}
        \left(\frac{\theta_i}{1}\right)^2 \left(\frac{m_a}{10^{-7}~\text{eV}}\right)^{-\frac32} \left(\frac{\beta}{10^{-13}}\right)^{-1} \left(\frac{\Min}{10^8~\text{g}}\right)^{-1} & \text{for } \Tosc \geq \Teq\,,\\
        \left(\frac{\theta_i}{1}\right)^2 \left(\frac{m_a}{10^{-8}~\text{eV}}\right)^{-2} \left(\frac{\Min}{10^8~\text{g}}\right)^{-\frac32} & \text{for } \Teq \geq \Tosc  \geq \Tc\,,\\
        \left(\frac{\theta_i}{1}\right)^2 \left(\frac{m_a}{10^{-8}~\text{eV}}\right)^{-2} \left(\frac{\Min}{10^8~\text{g}}\right)^{-\frac32} & \text{for } \Tc \geq \Tosc \geq \Tev\,,\\
        \left(\frac{\theta_i}{10^{-3}}\right)^2 \left(\frac{m_a}{m_a^\text{QCD}}\right)^{-\frac32} & \text{for } \Tev \geq \Tosc\,.
    \end{dcases}
\end{equation}

%%%%%%%%%%%%%%%%%%%%%%%%%%%%%%%%%%%%%%%%%%%%%%%%%%%%%%
\begin{figure}
    \def\sepf{0.57}
	\centering
	\includegraphics[scale=\sepf]{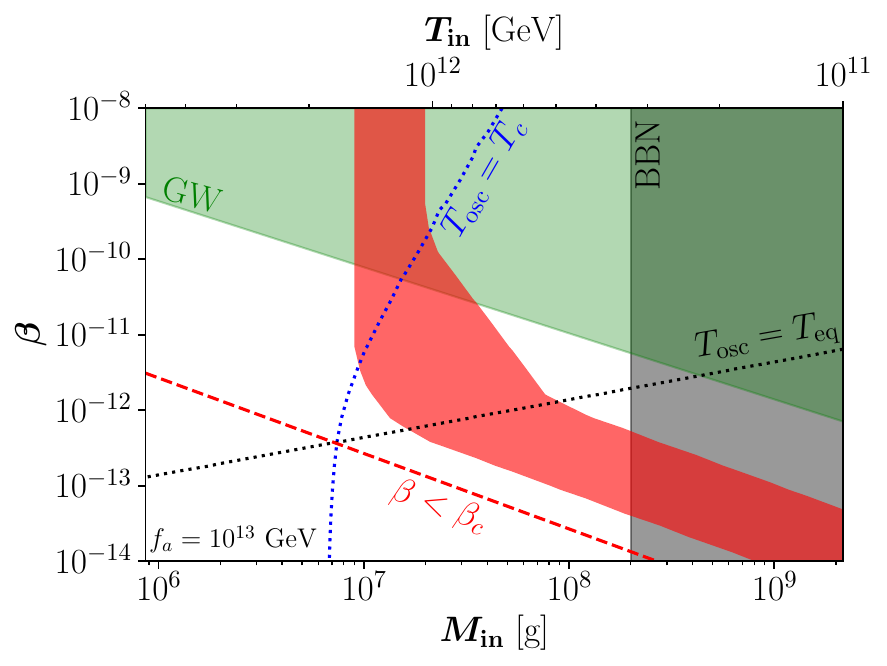}
	\includegraphics[scale=\sepf]{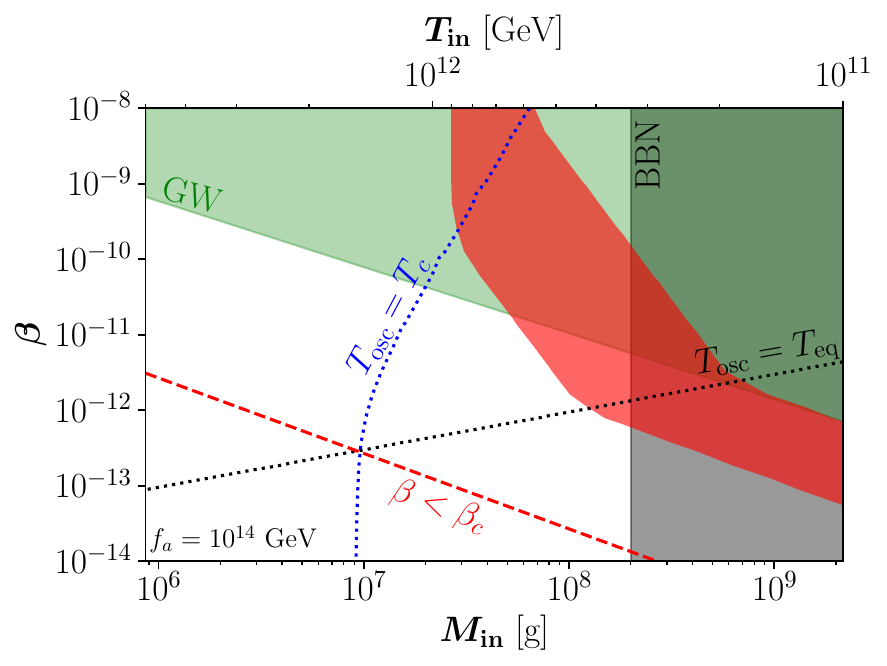}
    \caption{Parameter space generating the whole observed DM abundance for $f_a = 10^{13}$~GeV in the upper panel (which corresponds to $m_a \simeq 6 \times 10^{-7}$~eV) and $f_a = 10^{14}$~GeV in the lower panel (which corresponds to $m_a \simeq 6 \times 10^{-8}$~eV), taking $0.5 \leq \theta_i \leq \pi/\sqrt{3}$.
    }
	\label{fig:Min_beta}
\end{figure} 
%%%%%%%%%%%%%%%%%%%%%%%%%%%%%%%%%%%%%%%%%%%%%%%%%%%%%%
The red band in Fig.~\ref{fig:Min_beta} shows the parameter space generating the whole observed QCD axion DM abundance for $f_a = 10^{13}$~GeV in the upper panel (that  corresponds to $m_a \simeq 6 \times 10^{-7}$~eV) and $f_a = 10^{14}$~GeV in the lower panel (that  corresponds to $m_a \simeq 6 \times 10^{-8}$~eV), in the plane $[\Min,\, \beta]$, for $0.5 \leq \theta_i \leq \pi/\sqrt{3}$.
The region on the right (left) produces a DM underabundance (overabundance).
The colored bands are in tension either with BBN $\Tev \lesssim 4$~MeV (gray) or with GW, i.e. Eq.~\eqref{eq:GW} (green).
Below the red dashed line $\beta < \beta_c$ PBH energy density is always subdominant with respect to radiation, and therefore one has a standard cosmological history.
On the contrary, above that line $\beta > \beta_c$, a nonstandard cosmological expansion triggered by PBHs occurs.
The dotted lines correspond to $\Tosc = \Tc$ and $\Tosc = \Teq$, showing the borders of the regions described previously. 
This figure with $f_a = 10^{13}$~GeV (which corresponds to $m_a \simeq 6 \times 10^{-7}$~eV), can be understood with the first three cases shown in Eq.~\eqref{eq:DM}.

The red band in Fig.~\ref{fig:ma_theta} shows the misalignment angle required to reproduce to whole observed axion DM abundance for the case with PBH domination.
The lower border of the band (i.e. the red thick line) corresponds to the standard cosmological scenario.
The thickness of the band brackets all possible PBH scenarios compatible with BBN ($\Tev > 4$~MeV) and GW (i.e., Eq.~\eqref{eq:GW}).
PBHs enlarge the standard axion DM window, from $1.6 \times 10^{-6}~$eV $\lesssim m_a \lesssim 1.4 \times 10^{-5}$~eV in the standard case, to $2 \times 10^{-8}~$eV $\lesssim m_a \lesssim 1.4 \times 10^{-5}$~eV, for $0.5 < \theta_i < \pi/\sqrt{3}$, allowing to explore lighter axions.
It is worth emphasizing that this is mainly due to the effect of the entropy injection of the PBH Hawking evaporation.
Finally, the knee in Fig.~\ref{fig:ma_theta} around $m_a \simeq 10^{-10}$~eV is due to the temperature dependence of the axion mass.

%%%%%%%%%%%%%%%%%%%%%%%%%%%%%%%%%%%%%%%%%%%%%%%%%%%%%%%%%%
\section{Conclusions} \label{conclusion}
%%%%%%%%%%%%%%%%%%%%%%%%%%%%%%%%%%%%%%%%%%%%%%%%%%%%%%%%%%
In this paper we  studied  the production of QCD axions as dark matter (DM) candidates, in the scenario where primordial black holes (PBHs) dominate over the energy density of the universe, triggering a nonstandard cosmological epoch.

We focused on PBHs with masses below $\sim 2 \times 10^8$~g fully evaporate before the onset of Big Bang nucleosynthesis. PBHs modify the standard axion production in different ways: Firstly, axions are inevitably generated by Hawking radiation of PBH evaporation, however, being ultra-relativistic at production, they will not correspond to cold DM but to dark radiation with $\Delta N_\text{eff} \simeq 0.04$.
Interestingly, this contribution is within the projected reach of the future CMB Stage 4 experiment, and could relax the tension between late and early-time Hubble determinations.
Secondly, the oscillation temperature of axions reduces if there is a PBH dominated era, which leads to increase the allowed axion mass range.
Finally, the PBH Hawking evaporation injects entropy to the SM, diluting the axion relic abundance originally produced by the misalignment mechanism.  Considering the bound on the initial mass and abundance of PBHs, we find the the overall effect is that the parameter space with ultralight axions is opened up thanks to  the entropy injection due to the PBH Hawking evaporation, reaching $2 \times 10^{-8}$~eV $\lesssim m_a \lesssim 1.4 \times 10^{-5}$~eV, for a misalignment angle $0.5 < \theta_i < \pi/\sqrt{3}$. This is equivalent to the axion decay constant in the range $10^{11}~\text{GeV}\lesssim f_{a}\lesssim 10^{14}$~GeV.

It is interesting to note that axions typically couple to photons in many scenarios, like in the KSVZ~\cite{Kim:1979if, Shifman:1979if} or the DFSZ~\cite{Zhitnitsky:1980tq, Dine:1981rt} models.
In that case, the light axion window favored by a PBHs dominated epoch will be in the reach of future detectors like ABRACADABRA~\cite{Kahn:2016aff, Ouellet:2018beu}, KLASH~\cite{Alesini:2017ifp, Alesini:2019nzq}, and the next generation of ADMX~\cite{ADMX:2009iij, ADMX:2019uok}. 
Axion DM may therefore be a good probe of the history of the universe before Big Bang nucleosynthesis.

%%%%%%%%%%%%%%%%%%%%%%%%%%%%%%%%%%%%%%%%%%%
\section*{Acknowledgments}
NB received funding from the Spanish FEDER/MCIU-AEI under grant FPA2017-84543-P, and the Patrimonio Autónomo - Fondo Nacional de Financiamiento para la Ciencia, la Tecnología y la Innovación Francisco José de Caldas (MinCiencias - Colombia) grant 80740-465-2020.
F.H. thanks the support by the  project ``New Theoretical Tools for Axion Cosmology'' under the Supporting TAlent in ReSearch@University of Padova (STARS@UNIPD) and Istituto Nazionale di Fisica Nucleare (INFN) through the Theoretical Astroparticle Physics (TAsP) project.
This project has received funding /support from the European Union's Horizon 2020 research and innovation programme under the Marie Skłodowska-Curie grant agreement No 860881-HIDDeN. 

%%%%%%%%%%%%%%%%%%%%%%%%%%%%%%%%%%%%%%%%%%
\appendix
\section{Details of Computations}
\label{appa}
Here we present details for the derivation of the complex expression of third line of Eq.~\eqref{eq:H}. In the regime $\Tc \geq T \geq \Tev$, the entropy injection effects of PBHs evaporation dominate over the SM radiation, thus one could neglect the second term of Eq.~\eqref{boltz2}.
Plugging Eq.~\eqref{eq:dMdt} into the right-hand side of Eq.~\eqref{boltz2} and using the scale factor as variable, one has:
\begin{equation} \label{eq:sol_rhoR}
    \rR(a) \simeq \left[\left(\frac{a}{a_{\text{ev}}}\right)^{-3/2} -1 \right]   H(a_{\text{ev}})\, \left(\frac{\pi\gs(\Tbh)}{240}\frac{M_P^6}{\Mbh^3} \right) + \rR(a_{\text{ev}})\,.
\end{equation}
Since in the PBHs dominated phase, $H(a) \propto a^{-\frac{3}{2}}$, and note that   $H(a_{\text{ev}}) \simeq H_R(a_{\text{ev}})$, one has
\begin{align}
 H(a) & \simeq \left(\frac{a}{a_{\text{ev}}}\right)^{-3/2} H_R(a_{\text{ev}}) = \frac{\rR(a) -\rR(a_{\text{ev}}) +  H_R(a_{\text{ev}}) \, \left(\frac{\pi\gs(\Tbh)}{240}\frac{M_P^6}{\Mbh^3} \right)  }{ \left(\frac{\pi\gs(\Tbh)}{240}\frac{M_P^6}{\Mbh^3} \right) } \nonumber\\
&=  H_R(a_{\text{ev}}) + \frac{ 3M_P^2 \left( H^2_R(a) - H^2_R(a_{\text{ev}})\right) }{ \left(\frac{\pi\gs(\Tbh)}{240}\frac{M_P^6}{\Mbh^3} \right) } = H_R(a_{\text{ev}})  - \frac{720 \,\Mbh^3}{\pi\gs(\Tbh) M_P^4} \left(H^2_R(a_{\text{ev}}) - H^2_R(a)\right) \nonumber\\
    & =  H_R(a_{\text{ev}})  \left[1- \frac{720 \,\Mbh^3}{\pi\gs(\Tbh) M_P^4} \frac{H^2_R(a_{\text{ev}}) - H^2_R(a) }{ H(a_{\text{ev}}) } \right]\,,
\end{align}
 where we have used Eq.~\eqref{eq:sol_rhoR} in the second step, and therefore one gets the third line of Eq.~\eqref{eq:H}.
 
 Additionally, a couple of comments are in order concerning the calculation of the first line of Eq.~\eqref{eq:sratio}. First, let us consider the regime with $\Teq \geq T \geq \Tc$,  one has
\begin{align}
	\frac{S(T)}{S(\Tev)} & = \left( \frac{ \gss(T) }{\gss(\Tev)}  \right) \left( \frac{T}{\Tev}   \right)^3  \left( \frac{a(T)}{a(\Tev)}  \right)^3 = \left( \frac{ \gss(T) }{\gss(\Tev)}  \right) \left( \frac{T}{\Tev}   \right)^3 \left(\frac{H(T)}{H(\Tev)}\right)^{-2}  \nonumber\\
	& = \left( \frac{ \gss(T) }{\gss(\Tev)}  \right) \left( \frac{T}{\Tev}   \right)^3 \left(\frac{	
	H_R(\Teq) \left[\frac{\gss(T)}{\gss(\Teq)} \left(\frac{T}{\Teq}\right)^3\right]^{1/2}}{H(\Tev)}\right)^{-2} \nonumber \\
	& = \left( \frac{ \gss(T) }{\gss(\Tev)}  \right) \left( \frac{T}{\Tev}   \right)^3 \left(\frac{ H^2(\Tev) }{	  H_R^2(\Teq) \left[\frac{\gss(T)}{\gss(\Teq)} \left(\frac{T}{\Teq}\right)^3\right]}\right) \nonumber\\
	& =  \left( \frac{ \gss(T) }{\gss(\Tev)}  \right) \left( \frac{T}{\Tev}   \right)^3 \left(\frac{ \gs(\Tev)  \Tev^4}{	  \gs(\Teq)  \Teq^4 \left[\frac{\gss(T)}{\gss(\Teq)} \left(\frac{T}{\Teq}\right)^3\right]}\right) \nonumber \\
	& =   \frac{\gss(\Teq)}{\gss(\Tev)}\, \frac{\gs(\Tev)}{\gs(\Teq)}\, \frac{\Tev}{\Teq} \,.
\end{align}
Then for the case $T \geq \Teq$, since the entropy is approximately conserved,  one has 
\begin{align}
\frac{S(T)}{S(\Tev)} & \simeq \frac{S(\Teq)}{S(\Tev)} = \left( \frac{ \gss(\Teq) }{\gss(\Tev)}  \right) \left( \frac{\Teq}{\Tev}   \right)^3  \left( \frac{a_{\text{eq}}}{a_{\text{ev}}}  \right)^3 \nonumber \\
	& = \left( \frac{ \gss(\Teq) }{\gss(\Tev)}  \right) \left( \frac{\Teq}{\Tev}   \right)^3   \left(\frac{H(\Teq)}{H(\Tev)}\right)^{-2}  \simeq \frac{\gss(\Teq)}{\gss(\Tev)}\, \frac{\gs(\Tev)}{\gs(\Teq)}\, \frac{\Tev}{\Teq}\,,
\end{align}
where  from the second to third step we have used the fact that $H\propto a^{-3/2}$ in the matter domination phase and for the last step we have considered $H^2 \propto \gs(T)\, T^4$ at both  $\Tev$ and $\Teq$ boundaries. Thereafter, for $ T \geq \Tc$, one has an entropy injection given by first line of Eq.~\eqref{eq:sratio}.

\bibliographystyle{JHEP}
\bibliography{biblio}

\end{document}